\documentclass[hyper]{JHEP} 

\usepackage{epsfig}

\newcommand\fverb{\setbox\pippobox=\hbox\bgroup\verb}

\newcommand\fverbdo{\egroup\medskip\noindent%

			\fbox{\unhbox\pippobox}\ }

\newcommand\fverbit{\egroup\item[\fbox{\unhbox\pippobox}]}

\newbox\pippobox

\title{Remark about Non-BPS Dp-Brane 
at the Tachyon Vacuum 
Moving in Curved 
Background}

\author{by J. Kluso\v{n}\\

	 Department of Theoretical Physics and Astrophysics\\

                   Faculty of Science, Masaryk University\\

Kotl\'{a}\v{r}sk\'{a} 2, 611 37, Brno\\

Czech Republic\\

	E-mail: \email{klu@physics.muni.cz}}

\preprint{\hepth{0504062}}

\abstract{This paper is devoted to the study
of the dynamics of a non-BPS Dp-brane 
at the tachyon vacuum that moves
 in the  curved background.}
\keywords{D-branes}

\def\bx{\mathbf{x}}

\def\bA{\mathbf{A}}

\def\mH{\mathcal{H}}

\def\mE{\mathcal{E}}
\def\mK{\mathcal{K}}

\begin{document}
\section{Introduction}\label{first}
One of the most interesting problems
in string theory is the study of the time
dependent process. Even if this
problem is far from to be solved in 
the full generality one can find
many examples where we can obtain
some interesting results. The most
celebrated problem is the time
dependent tachyon condensation
in the open string theory
\footnote{For recent review and 
extensive list of references,
see \cite{Sen:2004nf}.}.  
Another  example of the time
dependent process  is the study of
the motion of the probe D-brane 
in given supergravity background. 
It turns out that the dynamics of
such a probe has a lot of common
with the time dependent tachyon 
condensation \cite{Kutasov:2004dj}
\footnote{Similar problems have
been discussed in
\cite{Thomas:2005fw,Kluson:2005jr,Huang:2005rd,
Thomas:2005am,Chen:2005wm,Kluson:2005qx,
Nakayama:2004ge,Chen:2004vw,Thomas:2004cd,
Bak:2004tp,Kluson:2004yk,Kluson:2004xc,
Saremi:2004yd,Kutasov:2004ct,Sahakyan:2004cq,
Ghodsi:2004wn,Panigrahi:2004qr,Yavartanoo:2004wb}.}.
In our previous works 
\cite{Kluson:2005jr,Kluson:2005qx,
Kluson:2004yk,Kluson:2004xc}
we have studied
the dynamics of a non-BPS Dp-brane
in the Dk-brane and in NS5-brane
background
in the effective field theory 
description. 
We have shown  that generally, when
we take the time dependent tachyon
into account, it is very difficult to
obtain  an exact time dependence of
the tachyon and radion mode. On the
other hand  we  argued in
\cite{Kluson:2005jr}, where we
 studied the properties
of the
worldvolume theory of BPS D-branes
and non-BPS Dp-branes in the near
horizon limit of  $N$ Dk-branes
or NS5-branes,
that the problem simplifies considerably
in case when  the
tachyon reaches its homogeneous 
 vacuum value $T_{min}$ that
is defined as $V(T_{min})=0
 \ , \partial_iT_{min}=0$ 
where $V(T)$ is a  tachyon potential.
Since the analysis in \cite{Kluson:2005jr}
was performed in the near horizon
region of given background configuration
of D-branes one can ask the question
how this description changes when 
we do not restrict to this particular
situation.  This paper is 
then  devoted 
to the study of the situation
when  
the non-BPS Dp-brane at the
tachyon vacuum  moves 
in  general spatial dependent background. 

An analysis of the properties of the
DBI non-BPS tachyon effective action 
at the tachyon vacuum 
was previously performed in  
\cite{Gibbons:2000hf,
Sen:2000kd,Yee:2004ec,Kwon:2003qn,Gibbons:2002tv,
Sen:2002qa}.
However this analysis was mainly focused
on the problem of the space-time filling non-BPS
Dp-brane. Our goal on the other hand
is to study the  dynamics of the
 non-BPS Dp-brane where the worldvolume
tachyon reaches its minimum 
and when this Dp-brane 
 is embedded in a
general background. 

As it is  believed  the final
state of the D-brane decay 
comprises  the dust of massive closed strings 
known as a  tachyon matter
  \cite{Sen:2002in,Sen:2002an,Lambert:2003zr}.
Another interesting aspect of the
low energy theory is found in the
sector with net electric flux that 
carries fundamental string charges.  
Generally, when the D-brane decays,
the classical solution of the system is 
characterised as a two component fluid 
system: One is preasurless electric
flux lines, known as string fluid, while
the other is a tachyon matter
\cite{Mukhopadhyay:2002en,Nagami:2003mr,
Sen:2003iv,Rey:2003zj,Rey:2003xs}.
As we claimed above the string
fluid and tachyon matter must have
a natural interpretation via closed
string states.  In fact, it was
shown that string fluid reproduces
the classical behaviour of fundamental
string remarkably well. Dynamics of
such a configuration has been shown 
to be exactly that of  Nambu-Goto
string
\cite{Gibbons:2000hf,
Sen:2000kd}. Natural construction
from this however, hampered by 
the degeneracy of the string fluid.

More recently the macroscopic 
interpretation for the combined
system of string fluid and tachyon
matter was proposed in 
\cite{Yee:2004ec,Sen:2003bc}. The basic
idea was to consider
a macroscopic number of long fundamental
strings lined up along one particular
directions and turn on oscillators
along each of these strings.
The proposed map is to identify energy
of electric flux lines as coming from
the winding mode part of the fundamental
strings, while attributing the tachyon matter
energy to oscillator part. 

While the
analysis performed in \cite{Yee:2004ec,Sen:2003bc}
is very interesting and certainly deserves
generalisation to the Dp-brane moving
in general background (We hope
to return to this problem in future) the
goal of this paper is more modest. 
As it is clear from the analysis 
given in \cite{Yee:2004ec,Sen:2003bc}
the crucial point in the mapping
 the string fluid and 
the tachyon matter
to the fundamental strings degrees
of freedom  is an
existence of the nonzero electric flux. 
On the other hand we know that the
tachyon condensation also occurs when
the electric flux is zero  and the resulting
configuration should correspond to
the gas of massive closed strings
\cite{Lambert:2003zr}. 
Due to the remarkable success of
the tachyon effective action in the
description of the open string tachyon
condensation one could hope that 
the classical effective field theory
analysis should be able to capture
some aspects of the closed strings
a non-BPS Dp-brane decays into. 
We will see that this is indeed
the case. 
 More precisely,
in section (\ref{second}) we will 
solve the equation of motion for
the non-BPS Dp-brane at the tachyon
vacuum moving in the Dk-brane background
and we will argue that the solution
is the same as the collective
 motion of the gas of massless particles.
 Then in section
(\ref{third}) we will demonstrate
the equivalence between the 
homogeneous  tachyon condensation and
the  gas of massless particles
for  spacetime, where the
metric components are functions 
of coordinates transverse 
to Dp-brane, 
 following \cite{Sen:2000kd}.
As we will argue  in the conclusion
this result is in perfect agreement
with the open-closed string conjecture
presented in \cite{Sen:2003iv,Sen:2003xs}.
In  order to find the
solution corresponding to the macroscopic
fundamental string we will 
consider the solution with 
nonzero electric flux aligned
along one spatial direction
on the worldvolume of the
Dp-brane. We will show
in section (\ref{fourth})
that this solution can be
interpreted as a gas of the
macroscopic strings stretched
along this direction that
move in given supergravity
background. 
Then the dynamics 
 of a non-BPS Dp-brane
with nonzero electric flux
 that moves  in Dk-brane background
will be studied in 
section (\ref{fifth}).
In conclusion (\ref{sixth})
we outline our result and 
suggest possible extension
of this work.
\section{Hamiltonian formulation
of the  Non-BPS Dp-brane}\label{second}
As we claimed in the introduction 
the main goal of this paper 
 is to study the tachyon
effective action at the tachyon vacuum.
Even if the Lagrangian for  
a non-BPS Dp-brane in its
tachyon vacuum  vanishes
\cite{Sen:1999md,Kluson:2000iy,
Bergshoeff:2000dq,Garousi:2000tr,Kutasov:2003er},
the dynamics of this
configuration  is still nontrivial
\cite{Gibbons:2000hf,
Sen:2000kd,Yee:2004ec,Kwon:2003qn,Gibbons:2002tv,
Sen:2002qa} as follows from
the fact that the Hamiltonian for 
a non-BPS  Dp-brane at the tachyon
vacuum is nonzero.

More precisely, let us  
introduce   the 
 Hamiltonian
for a non-BPS Dp-brane that is moving
in  $9+1$ dimensional
background with the metric 
\begin{equation}\label{genm}
ds^2=-N^2dt^2+g_{ab}(dx^a+L^adt)
(dx^b+L^bdt)  \ , a,b=1,\dots,9
\end{equation}
and with  the spatial dependent
dilaton \footnote{In this paper we will
consider the case when the metric
and dilaton are functions of the
coordinates transverse to Dp-brane
worldvolume. This restriction is
relevant for the study of the probe
non-BPS Dp-brane in the Dk-brane
background.}.
 Let us now
 consider the non-BPS action in the form
\begin{equation}\label{aclag}
S=-\int d^{p+1}\xi 
e^{-\Phi}V(T)\sqrt{-\det \bA} \ ,
\end{equation}
where
\begin{equation}
\bA_{\mu\nu}=G_{MN}\partial_{\mu} X^M
\partial_{\nu} X^N+F_{\mu\nu}+W(T)
\partial_{\mu}T\partial_{\nu}T  \ ,
\end{equation}
where $M,N=0,1,\dots,9$ and where
$V(T),W(T)$ are functions of $T$ that
vanish for $T_{\min}= \pm \infty$.  
Let us fix the gauge by 
$\xi^\mu=x^{\mu} \ , \mu=0,1,\dots,p$.
In what follows we will also use
the notation $\bx=(x^1,\dots,x^p)$.  
With the metric (\ref{genm}) the
components of the 
matrix  $\bA$ take the form
\begin{eqnarray}
\bA_{00}=-N^2+g_{ij}L^iL^j+g_{IJ}
\partial_0X^I\partial_0X^J+W(\partial_0T)^2 
\nonumber \\
\bA_{0i}\equiv E^+_i=
g_{ij}L^j+g_{IJ}\partial_0X^I
\partial_iX^J+
F_{0i}+W\partial_0 T\partial_i T 
\nonumber \\
\bA_{i0}\equiv -E^-_i=
g_{i0}+g_{ij}L^j+g_{IJ}\partial_iX^I
\partial_0X^J
-F_{0i}+W\partial_i T\partial_0 T \nonumber \\
\bA_{ij}=g_{ij}+g_{IJ}\partial_iX^I
\partial_jX^J+F_{ij}+W\partial_iT\partial_jT \ , 
\nonumber \\
\end{eqnarray}
where $i,j=1,\dots,p$ and $I,J=p+1,\dots,9$.  
Then we can write
\begin{equation}
\det \bA=
\bA_{00}\det \bA_{ij}+E^+_iD_{ij}E^-_j
 \ , 
D_{ij}=(-1)^{i+j}\triangle_{ji} \ , 
\end{equation}
where $\triangle_{ji}$ is the determinant
of the matrix with j-th row and
i-th column omitted. 
From (\ref{aclag}) we obtain the
canonical  momenta as
\begin{eqnarray}
\pi^i=\frac{\delta \mathcal{L}}
{\delta \partial_0 A_i}=
\frac{Ve^{-\Phi}}{\sqrt{-\det\bA}}
\frac{E^+_jD_{ji}+D_{ij}E^-_j}{2} \ ,
\nonumber \\
\pi_T=
\frac{\delta\mathcal{L}}
{\delta \partial_0T}=\frac{e^{-\Phi}
VW}{\sqrt{-\det\bA}}
\left(\dot{T}\det \bA_{ij}
-\frac{E^+_jD_{ji}\partial_iT-
\partial_iTD_{ij}E_j^-}{2}\right) \ ,
\nonumber \\
p_I=\frac{\delta\mathcal{L}}{\delta
\partial_0X^I}=\frac{e^{-\Phi}V}
{\sqrt{-\det\bA}}
\left(g_{IJ}\partial_0X^J\det\bA_{ij}
-\frac{E^+_jD_{ji}g_{IJ}\partial_iX^J
+g_{JI}\partial_iX^JD_{ij}E^-_j}{2}
\right)
\nonumber \\
\end{eqnarray}
Note also that $\pi^i$  satisfies
the Gauss law constraint $\partial_i\pi^i=0$. 
The Hamiltonian density 
is then obtained following
Legendre transformation
\begin{equation}
\mathcal{H}(\bx)=
\pi^iE_i+\pi_T\dot{T}+p_I\dot{X}^I
-\mathcal{L} 
 \ .
\end{equation}
After some length and tedious
 algebra 
we obtain the Hamiltonian density
as a function of canonical
variables 
in the form 
\begin{eqnarray}\label{hdenge}
\mathcal{H}=N
\sqrt{\mK}
-\pi^i F_{ij}L^j
-p_KL^K
+(\pi_T\partial_i T+p_K\partial_i X^K)L^i \ ,
\nonumber \\ 
\mK=\pi^i g_{ij}
\pi^j+W^{-1}
\pi_T^2+p_Ig^{IJ}p_J+
b_ig^{ij}b_j+
\nonumber  \\
+(\pi^i \partial_i T)^2
+(\pi^i \partial_i X^K)g_{KL}
(\pi^j \partial_j X^L)+
e^{-2\Phi}V^2\det \bA_{ij} \  , 
\nonumber \\
b_i=F_{ik}\pi^k+\pi_T\partial_i T+
\partial_i X^Kp_K \ . 
\nonumber \\
\end{eqnarray}
The form of the
Hamiltonian density (\ref{hdenge})
 considerably simplifies  
in situation when the
tachyon reaches its global minimum
($V(T_{min})=W(T_{min})=0$)
 and also when 
 its spatial derivatives are equal to
 zero: $\partial_iT=0$. 
This state is interpreted
as a final state of the unstable
Dp-brane decay that does not
contain any propagating open string
degrees of freedom.
 On the other hand
we  see that even in this
case there is still nontrivial dynamics
as follows from the form of the
Hamiltonian density (\ref{hdenge}).

To see this more clearly
 we begin  with 
an explicit 
example  of an unstable
 Dp-brane in its
tachyon vacuum that 
 moves in the background
of  $N$  coincident 
 Dk-branes.
The metric, the dilaton $(\Phi)$, 
and the R-R
field (C) for  a system of $N$ coincident 
Dk-branes is given by
\begin{eqnarray}\label{Dkbac}
g_{\alpha \beta}=H_k^{-\frac{1}{2}}\eta_{\alpha\beta}
\ , g_{mn}=H_k^{\frac{1}{2}}\delta_{mn} \ ,
(\alpha, \beta=0,1,\dots,k \ , 
m, n=k+1,\dots,9) \ , \nonumber \\
e^{2\Phi}=H_k^{\frac{3-k}{2}} \ , 
C_{0\dots k}=H_k^{-1} \ , H_k=1+\frac{\lambda}
{r^{7-k}} 
\ ,
\lambda=Ng_sl_s^{7-k}  \ , 
 \nonumber \\
\end{eqnarray}
where $H_k$ is a harmonic function of $N$ 
Dk-branes satisfying the Green 
function equation in
the transverse space. 
We will consider
a non-BPS Dp-brane with $p<k$ 
that is inserted
in the background (\ref{Dkbac}) with its
spatial section   stretched
in directions $(x^1,\dots,x^p)$. 
For zero  electric flux and  
for tachyon equal to $T_m$ 
the Hamiltonian density 
(\ref{hdenge}) takes the form
\begin{eqnarray}\label{hamdeni}
\mathcal{H}=N\sqrt{p_Ig^{IJ}p_J+
\partial_i X^Kp_K
g^{ij}\partial_j X^Lp_L
}
=N\sqrt{\mathcal{K}(\bx)}  \ .
\nonumber \\
\end{eqnarray}
Using  (\ref{hamdeni}) 
the canonical  equations of motion 
take the form 
\begin{eqnarray}\label{eqx}
\partial_0 X^K(\bx)=
\frac{\delta H}{\delta p_K(\bx)}=
N\frac{g^{KL}p_L+\partial_iX^K
g^{ij}\partial_jX^Lp_L}{
\sqrt{\mathcal{K}(\bx)}}
\nonumber \\
\end{eqnarray}
and 
\begin{eqnarray}\label{eqp}
\partial_0p_K(\bx)=-\frac{\delta H}
{\delta X^K(\bx)}=
-\frac{\delta N}{\delta X^K(\bx)}\sqrt{\mathcal{K}}
-\nonumber \\
-\frac{1}{2\sqrt{\mathcal{K}}}
\left(\frac{\delta g^{IJ}}{\delta X^K}
p_Ip_J+\partial_iX^IP_I\frac{\delta
g^{ij}}{\delta X^K}\partial_j X^JP_J
\right)
+\partial_i\left[\frac{NP_Kg^{ij}\partial_j
X^Lp_L}{\sqrt{\mathcal{K}}}\right] \ ,
\nonumber \\
\end{eqnarray}
where $N=\sqrt{-g_{00}}, \ g_{ij} \ , 
g_{IJ}$ and $\Phi$ are given
in (\ref{Dkbac}).

To further simplify the problem we 
 restrict ourselves
to the case of homogeneous modes on
the worldvolume of non-BPS Dp-brane.
Then the equations of motions
(\ref{eqx})  take the  form 
\begin{eqnarray}\label{eqxh}
\partial_0 X^m=
\frac{p_m}{H_k^{3/4}
\sqrt{\mathcal{K}}}
 \ , \nonumber \\
\partial_0Y^u=
\frac{H_k^{1/4}p_u}{\sqrt{\mK}} \ ,
\nonumber \\
\end{eqnarray}
where  
  $Y^u \ ,u,v=p+1,\dots,k$
are worldvolume modes that characterise
the transverse position of Dp-brane that
 is parallel with the worldvolume of Dk-branes
and $X^m \ , m=k+1,\dots,9$ are worldvolume
modes that parametrise transverse positions both
to the Dk-branes and to Dp-brane. 
Thanks to the manifest rotation invariance in
transverse $R^{9-k}$ space we will restrict ourselves
to the motion in the $(x^8,x^9)$ plane where
we introduce the cylindrical
coordinates
\begin{equation}
X^8=R\cos\theta \ ,
X^9=R\sin\theta \ .
\end{equation}
Note also since  the Hamiltonian
does not explicitly depend on 
$Y^u$ and $\theta$ the corresponding
conjugate momenta $p_u \ , p_\theta$
are conserved. 
As a next step we  use 
the fact that
the energy density 
\begin{equation}
\mE=\sqrt{-g_{00}}\sqrt{\mK}
\end{equation}
is conserved and replace
$\mK$ with $\mE$ and also  express 
$p_R$ as a  function of $R$
and conserved quantities
$\mE,p_u,p_\theta$ 
\begin{equation}
p_R=\pm\sqrt{H_k}
\sqrt{\mE^2-p_u^2-\frac{p^2_\theta}{R^2H_k}
} \  .
\end{equation}
Then the equation of motion
(\ref{eqxh}) can be written as
\begin{eqnarray}\label{eqt}
\partial_0Y^u=\frac{p_u}{\mE} \ , 
\nonumber \\
\partial_0 \theta=\frac{
p_\theta}{R^2\sqrt{H_k}\mE} \ , \nonumber \\
\partial_0R=
\pm\frac{\sqrt{\mE^2-p_u^2-\frac{p^2_\theta}
{R^2H_k}}}{\sqrt{H_k}\mE} \ . \nonumber \\
\end{eqnarray}
In order to study the general
properties of the radial motion
of the probe non-BPS Dp-brane
we will present the similar
analysis as was performed in
\cite{Burgess:2003mm}. 
First of all, note that the
Hamiltonian density for the
background (\ref{Dkbac}) 
 takes the form
\begin{equation}
\mH=\sqrt{-g_{00}}
\sqrt{p_ug^{uv}p_v+p_rg^{rr}p_r
+p_\theta g^{\theta\theta}p_\theta}=
\sqrt{p_u^2+\frac{p_R^2}{H_k}+
\frac{p_\theta^2}{R^2H_k}} \ 
\end{equation} 
that implies that $\mH$ 
is an increasing function
of $p_R$ so that the allowed range
of $R$ for the classical motion
can be found by plotting 
the effective potential $V_{eff}(R)$
that is defined as
\begin{equation}\label{ved}
V_{eff}(R)=\mH(p_R=0)=
\sqrt{p_u^2+\frac{p_\theta^2}{R^2H_k}}
\end{equation}
 against
$R$ and finding those $R$ for
which $\mE\geq V_{eff}(R)$. 
The properties of $V_{eff}$ depend
on $H_k$ that is monotonically
decreasing function of $R$ with
the limit $H_k\rightarrow
\frac{\lambda}{R^{7-k}}$ for
$R\rightarrow 0$ and with 
$H_k\rightarrow 1$ for $R\rightarrow
\infty$. For  $p_\theta\neq 0$
we obtain following asymptotic behaviour
of the potential (\ref{ved}) for $R\rightarrow
0$
\begin{itemize}
\item {\bf k=6}

In this case we obtain
\begin{equation}
V_{eff}\rightarrow
\frac{|p_\theta|}{\sqrt{\lambda}
\sqrt{R}} 
\end{equation}
and hence for nonzero $p_\theta$ the
potential diverges at the origin.
\item {\bf k=5}

Now the potential in
the limit $R\rightarrow 0$ approaches
to
\begin{equation}\label{Vef}
V_{eff}=\sqrt{p_u^2+
\frac{p^2_\theta}{\lambda}} \ . 
\end{equation} 
\item $\mathbf{k<5}$

In this case the effective
potential takes the form 
\begin{equation}
V_{eff}\approx 
\sqrt{p_u^2+\frac{p^2_\theta 
R^{5-k}}{\lambda}}
\end{equation}
that again implies that potential
 approaches  the constant $\sqrt{p_u^2}$
in the limit $R\rightarrow 0$. 
\end{itemize}
On the other hand for $R\rightarrow
\infty$ we get 
\begin{equation}
V_{eff}\rightarrow \sqrt{p_\mu^2} \ . 
\end{equation}
More precisely, looking at the
form of the potential for 
$k=6,5$
it is easy to see that these potentials
are decreasing functions of $R$. 
On the other hand for $k<5$ it can
be shown that $V_{eff}$ has extremum
at 
\begin{equation}\label{locmax}
R_{max}=\left(\frac{\lambda(5-k)}
{2}\right)^{\frac{1}{7-k}} \ . 
\end{equation}
Collecting these results
we obtain  following pictures for
the dynamics of the
non-BPS Dp-brane in its tachyon vacuum
moving in Dk-brane background. 
In the first case we consider 
non-BPS Dp-brane that moves towards the
stack of $N$ Dk-branes from the 
asymptotic
infinity $R=\infty$ at $t=-\infty$.
It reaches its turning point at 
\begin{equation}
1-\frac{p_u^2}{\mE^2}-\frac{p_\theta^2}
{\mE^2 R^2_TH_k}=0 
\Rightarrow R_T^2+\frac{\lambda}{R_T^{5-k}}
=\frac{p^2_\theta}{\mE^2
\left(1-\frac{p^2_u}{\mE^2}\right)} \ , 
\end{equation}
 and  then it moves outwards.
On the other hand 
from the existence of local maxima 
(\ref{locmax}) for $k<5$ 
it is clear that 
 the Dp-brane can
be in bounded region near the stack
of $N$ Dk-branes. To see this more
precisely let us solve 
the third equation in (\ref{eqt}) 
in the limit $\frac{\lambda}{R^{7-k}}\gg 1$.
In this case we obtain following
equation
\begin{equation}\label{dRs}
\frac{dR}{
\sqrt{\left(1-\frac{p^2_u}{\mE^2}\right)
R^{7-k}-\frac{p^2_\theta}{\mE^2\lambda}
R^{2(6-k)}}}=\pm \frac{dt}{\sqrt{\lambda}}
\end{equation}
that has  the solution
\begin{equation}
R^{5-k}=\frac{\lambda(\mE^2-p_u^2)}
{p^2_\theta}\frac{1}{
1+\left(\mp\frac{\mE^2-p_u^2}{2\mE p_\theta}t+
\sqrt{R^{k-5}_0-1}\right)^2} \ . 
\end{equation}
We see that now Dp-brane leaves the
worldvolume of Dk-branes at $t=-\infty$
and moves outwards until 
its  turning point at
$\dot{R}=0$ and   then
it moves towards  
the stuck of Dk-branes
that it again reaches at $t=\infty$. The
precise analysis of the dynamics of
the Dp-brane in the region $\frac{\lambda}
{R^{7-k}}\gg 1$ was performed in
\cite{Kluson:2005jr} where more
details can be found. 

As it is clear from (\ref{Vef})
the effective potential  takes 
very simple form when $p_\theta=0$. 
In this case 
the differential equation for $R$ is
equal to
\begin{equation}\label{dfpu}
\dot{R}=\pm \frac{\sqrt{\mE^2-p_u^2}}
{\sqrt{H_k}\mE}
\end{equation}
that can be explicitly solved 
in terms of hypergeometric functions.
However in order to gain better
physical meaning of this
physical situation it is useful
to consider the case when $p_u=0$.
 Then the
  equation (\ref{dfpu}) can
be rewritten in more
suggestive form
\begin{equation}
 -H_k^{-1/2}dt^2+H_k^{1/2}dR^2=0
\end{equation}
that is an equation of the radial
geodesics in Dk-brane background.

In summary, we have found that
a non-BPS Dp-brane where the
tachyon reaches its vacuum value
moves
in the background of $N$ Dk-branes
 as a gas
of massless particles that are
confined to  the worldvolume of
the original Dp-brane. 
In the next section we will 
present more detailed arguments
that support validity of
  this correspondence.
\section{Non-BPS Dp-brane
at the Tachyon Vacuum 
as a Gas of Massless Particles}\label{third}
Let us consider  the 
 curved background with the
metric
\begin{equation}\label{met}
ds^2=-N^2dt^2+g_{ab}(dx^a+L^adt)
(dx^b+L^bdt)  \ , a,b=1,\dots,9 \ , 
\end{equation}
where we presume that $N,L^a,g_{ab}$ and
the dilaton $\Phi$ are functions
of the coordinates transverse
to Dp-brane.  
As we know from the previous
section the dynamics of the non-BPS Dp-brane
at the tachyon vacuum is governed
by the  Hamiltonian
 \begin{eqnarray}\label{hamdeng}
H=\int d\bx \mathcal{H} \ , 
\mathcal{H}=
N\sqrt{\mathcal{K}(\bx)}+p_K
\partial_i X^KL^i-p_KL^K  \ , \nonumber \\ 
\nonumber \\
\mK=p_Ig^{IJ}p_J+
\partial_i X^Kp_K
g^{ij}\partial_j X^Lp_L \ . \nonumber \\ 
\end{eqnarray}
It is now straightforward
to determine  the canonical equations 
of motions 
\begin{eqnarray}\label{eqpfo}
\partial_0 X^K(\bx)=
N\frac{g^{KL}p_L+\partial_iX^K
g^{ij}\partial_jX^Lp_L}{
\sqrt{\mathcal{K}(\bx)}}
+\partial_iX^KL^i -L^K
\nonumber \\
\end{eqnarray}
and 
\begin{eqnarray}\label{eqxfo}
\partial_0p_K(\bx)=
-\frac{\delta N}{\delta X^K(\bx)}
\sqrt{\mathcal{K}}
-\frac{1}{2\sqrt{\mathcal{K}}}
\left(\frac{\delta g^{IJ}}{\delta X^K}
p_Ip_J+\partial_iX^Kp_K
\frac{\delta g^{ij}}{\delta X^K}
\partial_j X^Lp_L\right)+
\nonumber \\
+\partial_i\left[\frac{Np_K
g^{ij}\partial_j
X^Lp_L}{\sqrt{\mathcal{K}}}\right]+
\partial_i[p_KL^i]-p_L\partial_iX^L
\frac{\delta 
L^i}{\delta X^K}+p_L\frac{\delta L^L}
{\delta X^K} \ . 
\nonumber \\
\end{eqnarray}
As we argued in the previous
section the Dp-brane at the tachyon vacuum
with zero electric flux 
has similar properties as a
homogeneous gas of the massless particles
embedded in the background of 
$N$ Dk-branes. Now we would like to
show that this correspondence holds
in more general situations. To see
this we
will closely follow very nice analysis
performed in  
 \cite{Sen:2000kd}. 

We begin with an 
 action for massive particle in general
spacetime 
\begin{equation}\label{actge}
S=-m\int d\tau \sqrt{-g_{MN}
\dot{Z^M}\dot{Z^N}}=-
m\int d\tau
\sqrt{\bA} \ ,
\end{equation}
where $\dot{Z}\equiv \frac{dZ}{d\tau}$ 
and where $Z^M$ are embedding coordinates
for massive particle.  
As a next step we fix the gauge in the form
$\tau=Z^0$ so that the
action (\ref{actge}) takes the form
\begin{eqnarray}\label{actgef}
S=-m\int d\tau
\sqrt{N^2-g_{st}L^sL^t-2g_{st}L^t\dot{Z}^s
-g_{st}\dot{Z}^s\dot{Z}^t}=
\nonumber \\
=-m\int d\tau
\sqrt{\bA} \ , s,t=1,\dots, 9 \ .
\nonumber \\  
\end{eqnarray}
Then the conjugate momenta 
 are
\begin{equation}
P_s=\frac{\delta S}{\delta
\dot{Z}^s}=
\frac{m\left(
g_{st}\dot{Z}^t+g_{st}L^t\right)}
{\sqrt{
\bA}} 
\end{equation}
and consequently the Hamiltonian
takes the form
\begin{eqnarray}
H=P_s\dot{Z}^s-L=
=N\sqrt{P_sg^{st}P_t+m^2}-P_sL^s  \ . 
\nonumber \\
\end{eqnarray}
Using the Hamiltonian formalism
 we can
take the limit $m\rightarrow 0$ 
and we obtain the Hamiltonian for
a massless particle moving
 in general
background
\begin{equation}
H=N
\sqrt{P_rg^{rs}P_s}-
P_sL^s \ . 
\end{equation}
Then the canonical  equations of motion 
for the massless particle take the
form 
\begin{eqnarray}\label{mpeqm}
\dot{Z}^s=\frac{\delta H}{\delta P_s}
=N\frac{g^{st}P_t}
{\sqrt{P_rg^{rs}P_s}}-L^s \ ,  \nonumber \\
\dot{P}_s=-\frac{\delta H}{\delta
Z^s}=
-\frac{\delta N}{\delta Z^s}
\sqrt{P_rg^{rt}P_t}-\frac{N}{2
\sqrt{P_rg^{rt}P_t}}
\frac{\delta g^{rt}}
{\delta Z^s}P_rP_t 
+P_r\frac{\delta L^r}{\delta Z^s} \ . 
\nonumber \\ 
\end{eqnarray}
Following  \cite{Sen:2000kd} 
we will now presume that
there exist 
solution of the 
equation of motion (\ref{mpeqm})
 given as
$Z^s(\tau),P_s(\tau)$. 
Consider 
then  the following field configuration
on the Dp-brane:
\begin{equation}\label{ansmp}
p_I(x^0,\dots,x^p)=
\left.P_I(\tau)
f(x^0,\dots,x^p)\right|_{\tau=x^0} \ , 
\end{equation}
where $f$ is an 
arbitrary
function of the variables
$(x^i-Z^i(\tau))$ for
$i=1,\dots,p$. Then it is clear
that
\begin{equation}
\left.
\left(\partial_0 f+\partial_i f \partial_\tau
 Z^i\right)\right|_{\tau=x^0}=0 \ . 
\end{equation}
We also demand that
$X^I$ obey 
\begin{equation}\label{xder}
\left.\left(\partial_iX^IP_I+P_i\right)
\right|_{\tau=x^0}=0 \ 
\end{equation}
but are otherwise unspecified.
Inserting the ansatz (\ref{ansmp})
into (\ref{hamdeng})
we obtain that the Hamiltonian
density 
takes the form 
\begin{eqnarray}\label{hhmp}
\mathcal{H}(x^0,\dots,x^p)
=\left(N(X)\sqrt{P_sg^{st}(X)P_t}-
P_sL^s\right)  
f(x^0,\dots,x^p)  \ .
\nonumber \\
\end{eqnarray}
We see that the expression
in the bracket has the form
of the Hamiltonian for the massless particle
where however the metric components
still  depend
on $X^I$ that are arbitrary
functions of
$t,\bx$. It turns out
however 
that in order to obey the
equation of motion for
general spacetime 
we should perform the
identification
\begin{equation}\label{idxz}
X^K(x^0,\dots,x^p)=Z^K(\tau) \ . 
\end{equation}
Then the equation of motion
(\ref{eqxfo})
can be written as
\begin{eqnarray}
\left(\partial_\tau P_K+
\frac{\delta N}
{\delta Z^K}\sqrt{
P_rg^{rt}P_t}
+\frac{N}{2\sqrt{P_rg^{rt}P_t}}
\left(\frac{\delta g^{IJ}}
{\delta Z^K}
P_IP_J+p_i
\frac{\delta g^{ij}}
{\delta Z^K}
p_j\right)
-P_L\frac{\delta L^L}{\delta Z^K}
\right)f
-\nonumber \\
-P_K\partial_if
\left(\partial_\tau Z^i
-\frac{Ng^{ij}P_j}
{\sqrt{P_rg^{rt}P_t}}
+L^i\right)\partial_i f=0 \ .
\nonumber \\
\end{eqnarray}
We see that this equation is obeyed
since the expressions
in the brackets are equal to zero
thanks to the fact that $Z^s,P_s$
obey the equations of motion 
(\ref{mpeqm}). 
On the 
other hand from
(\ref{xder}) and (\ref{idxz})
we get that  
$P_i=0$ and hence the configuration on
a non-BPS Dp-brane in the tachyon
vacuum  corresponds to the
motion of massless  particles 
that have  nonzero transverse 
momenta only. Then
the equation (\ref{eqxfo})
takes the form 
\begin{eqnarray}
\partial_0 X^K(\bx)=
\partial_\tau Z^K(\tau)=
N\frac{g^{KL}P_K}
{\sqrt{P_sg^{st}P_t}}-L^K
\nonumber \\
\end{eqnarray}
that is clearly obeyed since
$Z^K$ obeys (\ref{mpeqm}). 

The final question, and the
most difficult one, is regarded
to the form of the function $f(x^0,\dots,x^p)$.
We have seen that its form is not
determined from the Dp-brane
equations of motion. The most natural
choice is
\begin{equation}
f(x^0,\dots,x^p)=
\prod_{i=1}^p
\delta(x^i-Z^i(x^0)) \ . 
\end{equation}
As follows from (\ref{hhmp}) the
energy is localised along the
line $x^i=Z^i(x^0)$ for $i=1,\dots,p$. 
Using also the identification 
(\ref{idxz})  we see that in
the full $9+1$ dimensional spacetime
this solution describes the
worldline $x^s=Z^s(\tau)$ for
$s=1,\dots,9$.  In other words,
the Dp-brane worldvolume theory
contains a solution whose dynamics
is exactly that of massless particle
in $(9+1)$ dimensions. 

As in the case of NG string solution
given in \cite{Sen:2000kd} the freedom
of replacing the $\delta$ function
by an arbitrary function of 
$x^i-Z^i(\tau)$  is slightly unusual.
Very nice and detailed discussion
considering this issue was given in
\cite{Sen:2003bc}. According to this
paper the solution with arbitrary function
$f$ should be regarded as a system
of high density of massless particles,
or more precisely as a system of high
density of point-like solutions of
the closed string equations of motion. 
\section{Motion of Non-BPS Dp-brane
with nonzero electric flux}\label{fourth}
As we have seen in previous 
section the case
when the non-BPS Dp-brane in
the tachyon vacuum  moves
in the general background  
with zero electric flux 
can be interpreted as a 
motion of the gas of 
massless particles. In order
to find the solution of
the D-brane equations
of motion  having the
interpretation as a
 fundamental macroscopic string
we should rather consider the
case when we switch on
the electric  flux as well. 
In fact, let us again 
consider the  Hamiltonian for
a non-BPS Dp-brane 
at the tachyon vacuum
that moves 
in curved background 
\begin{eqnarray}\label{hdenf}
\mathcal{H}=
N\sqrt{\pi_i g^{ij}
\pi_j+p_Ig^{IJ}p_J+
b_ig^{ij}b_j
+(\pi^i \partial_i X^K)g_{KL}
(\pi^j \partial_j X^L)}+
\nonumber \\
+p_K\partial_iX^KL^i-
p_KL^K \ , 
\end{eqnarray}
where
\begin{equation}
b_i=F_{ij}\pi^j+
\partial_iX^Kp_K \ . 
\end{equation}
Note that $\pi^i$ 
also obey   the Gauss
law constraint
\begin{equation}\label{Glcg}
\partial_i\pi^i=0 \ . 
\end{equation}
Now canonical 
 equations of motion 
takes the form
\begin{equation}\label{aeq}
\partial_0A_i(\bx)=E_i(\bx)=
\frac{\delta H}
{\delta 
\pi^i(\bx)}
=\frac{N}{\sqrt{\mK}}
(g_{ij}\pi^j-F_{ik}g^{kj}b_j
+\partial_iX^Kg_{KL}(\pi^j
\partial_jX^L)) \ , 
\end{equation}
\begin{equation}\label{pieqg}
\partial_0\pi^i(\bx)=
-\frac{\delta H}{\delta A_i(\bx)}=
-\partial_j\left[\frac{N}
{\sqrt{\mK}}\left(\pi^jg^{ik}b_k
-\pi^ig^{jk}b_k\right)\right] \ ,
\end{equation}
\begin{equation}\label{xeq}
\partial_0X^I(\bx)=
\frac{\delta H}{\delta p_I(\bx)}=
\frac{N}
{\sqrt{\mK}}
\left(g^{IK}p_K+\partial_iXg^{ij}b_j\right)
+\partial_iX^KL^i
-L^K 
\ , 
\end{equation}
\begin{eqnarray}\label{Pieq}
\partial_0p_I(\bx)=
-\frac{\delta H}{\delta X^I(\bx)}=
\partial_i\left[\frac{N}
{\sqrt{\mK}}
\left(\pi^ig_{IK}\partial_jX^K\pi^j+
p_Ig^{ij}b_j\right)\right]+
\nonumber 
\\
+\frac{\delta N}{\delta X^I}
\sqrt{\mK}
-\frac{\sqrt{N}}{2\sqrt{\mK}}
\left(\pi^i\frac{\delta g_{ij}}{\delta
X^I}\pi^j-p_K\frac{\delta g^{KL}}
{\delta X^I}p_L-b_i\frac{\delta g^{ij}}
{\delta X^I}b_j-(\pi^i\partial_iX^K)
\frac{\delta g_{KL}}{\delta X^I}
(\pi^j\partial_jX^L)\right)+ \nonumber \\
+\partial_i[p_KL^i]
-p_L\partial_iX^L\frac{\delta L^i}
{\delta X^K}
+p_L\frac{\delta L^L}{\delta X^K} \ .
\nonumber \\ 
\end{eqnarray}
Following \cite{Sen:2000kd} we will
now try to find the solution
of the equation of motion given
above that can be interpreted as
 the fundamental string solution. 
To begin with let us consider 
the Nambu-Goto action for
fundamental  string 
\begin{equation}
S=-\int d\tau d\sigma
\sqrt{-\det G_{\alpha\beta}} \ , 
G_{\alpha\beta}=G_{MN}
\partial_\alpha Z^M\partial_\beta Z^N \ , 
\end{equation}
where $\alpha,\beta=\sigma,\tau$. 
We fix the gauge so that $Z^0=\tau,Z^1=\sigma$ 
so that
\begin{equation}
G_{\alpha\beta}=
g_{\alpha\beta}+
g_{st}\partial_\alpha Z^s
\partial_\beta Z^t \ , 
\end{equation}
where $s,t=2,\dots,9$. Then
the Hamiltonian takes the form
\begin{equation}
H_{NG}=\int d\sigma \mathcal{H}_{NG}(\sigma)\ ,
\end{equation}
where the Hamiltonian
density $\mathcal{H}_{NG}$ 
 is equal to
\begin{eqnarray}
\mathcal{H}_{NG}=N
\sqrt{\mK_{NG}}
-P_sL^s+P_s\partial_\alpha X^sL^\alpha \ ,
\nonumber \\
\mK_{NG}=
g_{\sigma\sigma}+
P_s g^{st}P_t+\partial_{\sigma}Z^s
P_sg^{\sigma\sigma}
\partial_\sigma Z^tP_t+
\partial_\sigma Z^s
\partial_\sigma Z^t g_{st}
\ . \nonumber \\
\end{eqnarray}
Now the equation of motion 
of the fundamental string take
the form
\begin{eqnarray}\label{eqzs}
\partial_\tau Z^s=\frac{\delta H_{NG}}{
\delta P_s}=
\frac{N\left(
g^{st}P_t+\partial_\sigma Z^s
g^{\sigma\sigma}\partial_\sigma Z^tP_t\right)}
{\sqrt{\mathcal{K}_{NG}}}-L_s
+\partial_\sigma X^sL^\sigma \ , 
\nonumber \\ 
\partial_\tau P_s=
-\frac{\delta H}{\delta Z^s}=
-\frac{\delta N}{\delta Z^s}
\sqrt{\mathcal{K}_{NG}}-\nonumber \\
-\frac{N}{2\sqrt{\mathcal{K}_{NG}}}
\left(\frac{\delta g_{\sigma\sigma}}
{\delta Z^s}+
P_r\frac{g^{rt}}{\delta Z^s}
P_t+\partial_\sigma Z^rP_r\frac{g^{\sigma\sigma}}
{\delta Z^s}\partial_\sigma Z^tP_t
+\frac{\delta g^{rt}}{\delta Z^s}
\partial_\sigma Z^r\partial_\sigma Z^t
\right) 
+\nonumber \\
+\partial_\sigma\left[
\frac{N\left(
P_s g^{\sigma\sigma}
\partial_\sigma Z^r P_r
+g_{st}\partial_\sigma Z^t\right)}
{\sqrt{\mathcal{K}_{NG}}}\right]
+P_t\frac{\delta L^t}{\delta X^s}
-P_t\partial_\sigma
 X^t\frac{\delta L^\sigma}
{\delta X^s}+
\partial_\sigma\left[
P_sL^{\sigma}\right] \ . 
\nonumber \\
\end{eqnarray}
For future use we also define
\begin{equation}
P=-\sum_{s=2}^9P_s
\partial_\sigma Z^s \ , 
Z^1(\tau,\sigma)=\sigma \ . 
\end{equation}
Let $Z^s(\tau,\sigma),
P_s(\tau,\sigma) \ , s=2,\dots,9 $ 
be the solutions
of the equation of motion 
(\ref{eqzs}). As was shown in 
\cite{Sen:2000kd} it
is natural to consider the following
field configuration on Dp-brane
\begin{eqnarray}\label{ansg}
\pi_i(x^0,\dots,x^p)=
\partial_\sigma
Z^i(\tau,\sigma)
f(x^0,\dots,x^p)|_{
(\tau,\sigma)=(x^0,x^1)} \ , \nonumber 
\\
p_I(x^0,\dots,x^p)=P_I(\tau,\sigma)
f(x^0,\dots,x^p)|_{(\tau,\sigma)
=(x^0,x^1)} \ , \nonumber \\
\end{eqnarray}
where $i=1,\dots,p$. 
Following \cite{Sen:2000kd} we
presume that $f(x^0,\dots,x^p)$ is
an arbitrary function of variables
$(x^m-Z^m(x^0,x^1))$ for $m=2,\dots,p$
and hence satisfies:
\begin{eqnarray}
\left.
\partial_\sigma Z^i
\partial_if\right|_{(\tau,\sigma)=
(x^0,x^1)}=0 \ , 
\left.\left(
\partial_0f+
\partial_i f \partial_\tau Z^i
\right)\right|
_{(\tau,\sigma)=
(x^0,x^1)}=0 \ .
\nonumber \\
\end{eqnarray}
We also presume that the fields
$X^I(x^0,\dots,x^p)$ and
 $F_{ij}(x^0,\dots,x^p)$
are subject following set of conditions:
\begin{eqnarray}
(\partial_\sigma Z^j\partial_jX^I
-\partial_\sigma Z^I)|_{(\tau,\sigma)=
(x^0,x^1)}=0 \ ,
\nonumber \\
(F_{ij}\partial_\sigma Z^j
+\partial_iX^IP_I+P_i)|_{(\tau,\sigma)=
(x^0,x^1)}=0 \ .
\nonumber \\
\end{eqnarray}
With this notation we can easily find that
\begin{eqnarray}
\pi^i\partial_iX^I(x^0,\dots,x^p)=
\partial_\sigma Z^I(\tau=x^0,
\sigma=x^1)f(x^0,\dots,x^p)  
\nonumber \\
b_i(x^0,\dots,x^p)=
-P_i(\tau=x^0,\sigma=x^1)
f(x^0,\dots,x^p) \ , \nonumber \\
\sqrt{\mK}(x^0,\dots,x^p)=
\sqrt{\mK_{NG}}(\tau=x^0,\sigma=x^1,X^I)
f(x^0,\dots,x^p) \ . 
\nonumber \\
\end{eqnarray}
We see that 
due to the nontrivial dependence of the
metric on transverse coordinates $X^I$ 
 the expression $\sqrt{\mK_{NG}}$ still
depends  on $X^I$. 
As in the case of  the particle-like 
solution studied in previous section
it is clear that in the curved spacetime
we should demand that
the  coordinates $X^I$
are related to $Z^I$ as: 
\begin{equation}\label{ecs}
X^I(x^0,x^1,x^m=Z^m(x^0,x^1))=
Z^I(x^0,x^1) \ . 
\end{equation}
This condition implies that 
\begin{equation}
\mathcal{H}(x^0,\dots,x^p)
=\mH_{NG}(\tau=x^0,\sigma=x^1)
f(x^0,\dots,x^p) \ . 
\end{equation}
Then we can show exactly as in
\cite{Sen:2000kd} 
 that the ansatz
(\ref{ansg}) together with
(\ref{ecs}) obeys the equations of
motion  
(\ref{aeq}),(\ref{pieqg}),
(\ref{xeq}),
(\ref{Pieq}) as well as the  
Gauss law constraint (\ref{Glcg}).
The interpretation of this solution
is the same as in the flat space
\cite{Sen:2003bc}. Firstly, the spatial
choice of the function $f$
\begin{equation}
f(x^0,\dots,x^p)=
\prod_{m=2}^p
\delta(x^m-Z^m(x^0,x^1))
\end{equation}
gives the solution that corresponds
to the stretched string in the $x^1$
direction that moves 
the  background (\ref{met}). On the other
hand the solutions with the general
form of the functions $f$ should
be interpreted as the configurations
of the high density of fundamental
strings moving in (\ref{met})
and 
that are confined to the worldvolume
of the original Dp-brane.
\section{Non-BPS Dp-brane at the
Tachyonic Vacuum with nonzero flux
in Dk-brane background}\label{fifth}
Let us again return to the 
spatial case of the
motion of the non-BPS Dp-brane
in its tachyon vacuum in the
Dk-brane background (\ref{Dkbac}).
As in  section (\ref{second})
we demand that all worldvolume
fields are homogeneous 
 $\partial_i X^I=0$ and the
electric flux has nonzero
component in the   $x^1$ direction
only:
\begin{equation}\label{gauge}
A_1=f(t) 
 \ , F_{ij}=0 \ . 
\end{equation}
For homogeneous fields and for the
gauge fields given in  (\ref{gauge})
we  get that $b_i=0$
and consequently $\mK$ in 
(\ref{hdenf}) is 
equal to
\begin{equation}
\mK=(\pi_1)^2H_k^{1/2}+
H_k^{-1/2}p_mp_m+H_k^{1/2}
p_up_u \ , 
\end{equation}
where  $p_m,m=k+1,\dots, 9$ are momenta
conjugate to coordinates $X^m$
transverse to Dk-brane and to Dp-brane
while $p_u,u=p+1,\dots, k$ are momenta
conjugate to coordinates $Y^u$
 transverse to 
Dp-brane but parallel to the Dk-brane.
With this ansatz 
the equation of motions 
(\ref{aeq}),(\ref{pieqg}),
(\ref{xeq}),
(\ref{Pieq}) take the form 
\begin{equation}\label{aeg}
\partial_0A_1(\bx)=E_1(\bx)=
\frac{\pi^1}{H_k^{3/4}\sqrt{\mK}}
  \ , 
\end{equation}
\begin{equation}\label{pieq} 
\partial_0\pi^i(\bx)=0 \ , 
\end{equation}
\begin{equation}\label{xmeq}
\partial_0X^m(\bx)=
\frac{p_m}{H_k^{3/4} \sqrt{\mK}} \ ,
\end{equation}
\begin{equation}\label{xieq}
\partial_0Y^u(\bx)=
\frac{p_uH_k^{1/4}}{ \sqrt{\mK}} \ .
\end{equation}
It is clear that the solution of (\ref{pieq})
consistent with the presumption that
all fields are homogeneous is the
constant electric flux $\pi_1=\Pi$. 
Note however that $E_1$ is time
dependent 
as follows from (\ref{aeg}) since
generally 
metric components depend on the coordinates
$X^m(t)$. 
To find the trajectory of  a non-BPS Dp-brane
we express $\sqrt{\mK}$ using the conserved
energy density as
\begin{equation}
\mathcal{E}=\sqrt{-g_{00}}
\sqrt{\mK}
\Rightarrow 
\sqrt{\mK}=\frac{\mathcal{E}}{\sqrt{-g_{00}}}
\end{equation}
so that
\begin{equation}
\partial_0 X^m=
\frac{p_m}{H_k\mathcal{E}} \ ,
\partial_0 Y^u=\frac{p_u}{\mathcal{E}} \ ,
E_1=\frac{\Pi}{H_k\mathcal{E}} \ .
\end{equation}
Since the Hamiltonian density
does not depend on  
$Y^u$ we get that $p_u=\mathrm{const}$. 
Using manifest rotation invariance
of the transverse $R^{9-k}$ space
we restrict ourselves to the
motion in $(x^8,x^9)$ plane 
where we also 
 introduce the $R$ and $\theta$
coordinates defined as 
\begin{equation}
X^8=R\cos\theta \ , 
X^9=R\sin\theta \ . 
\end{equation} 
Using the fact that
$p_\theta$ is conserved
we   
express $p_R$ from
 $\mE$  as 
\begin{eqnarray} 
p_R=\pm\sqrt{H_k}\sqrt{
\mE^2-\Pi^2-p_u^2-\frac{p^2_\theta}{R^2H_k}}
\nonumber \\
\end{eqnarray}
so that we  get
\begin{equation}\label{dotRs}
\dot{R}=\frac{p_R}{\sqrt{H_k}
\mE}=\pm\frac{\sqrt{
\mE^2-\Pi^2-p_u^2-
\frac{p^2_\theta}{R^2H_k}}}
{\sqrt{H_k}\mE} \ . 
\end{equation}
Since the equation  
(\ref{dotRs}) has the same form
as the equation (\ref{eqt}) 
(If we identify $\mE^2-\Pi^2$ in
(\ref{dotRs}) with $\mE^2$ in
(\ref{eqt}).) then the analysis
of the equation
(\ref{eqt}) 
performed in  section (\ref{second})
holds for (\ref{dotRs}) as well. 
Then we can interpret the
solution with   nonzero electric flux
$\Pi$  as a  solution
describing the motion of the
homogeneous gas  of the
macroscopic strings stretched
 in the $x^1$ direction 
that are confined to the worldvolume
of a non-BPS Dp-brane
and that
move in  Dk-brane background.

\section{Conclusion}\label{sixth}
We have studied the dynamics
of the non-BPS Dp-brane
at the tachyon vacuum and when
this Dp-brane  moves in the 
background  where metric
and dilaton are functions
of the coordinates transverse
to 
Dp-brane. We have shown that 
in case when there is not any electric
flux present on the worldvolume
 of this Dp-brane,
its dynamics is equivalent
to the dynamics of the homogeneous gas
of massless particles that are confined
on the worldvolume of the unstable Dp-brane.
At this place we should ask the
question how this  result 
is related to the analysis performed
 in \cite{Lambert:2003zr}
where it was shown that the end product
of the tachyon condensation should be
the gas of massive closed strings.
Relevant problem has been discussed
in \cite{Sen:2003xs}. According
to this paper there exists 
the spread of the energy density from
the plane of the brane due to the 
internal oscillation of the final state
of the closed strings. In the classical
limit we have delta function localised
D-brane and hence according to previous
remark the state 
of closed strings without oscillator
excitations. This however also implies
that the classical description of
such  closed strings is given
by   massless like solution
of the equation of motion when the modes
on the worldvolume of fundamental string
are not function of $\sigma$
\footnote{For recent review of
some aspects of classical string solutions,
see \cite{Tseytlin:2003ii}.}. 
In other words, the  classical result
given in this paper can be
considered as a manifestation
of the \emph{Open Closed Duality Conjecture}
proposed in \cite{Sen:2003iv}. 

In order to find macroscopic fundamental
string solutions we had to, 
as in the flat spacetime,  consider
the nonzero electric flux on the worldvolume
of the non-BPS Dp-brane. Then we have
shown that the dynamics of the unstable
D-brane at the tachyon vacuum with
the nonzero electric flux corresponds to
the dynamics of the gas of stretched
fundamental strings. 

In concussion, we would like to stress
that the results presented  in this paper
gives the modest
contribution to the study of the
tachyon condensation. On the other
hand we hope that they 
could be helpful for better
understanding of the general properties
of the tachyon condensation in curved
spacetime. 

{\bf Acknowledgement}

This work was supported by the
Czech Ministry of Education under Contract No.
MSM 0021622409.


\end{document}